\documentclass[journal]{journal}
\usepackage{amssymb}
\usepackage{graphicx}
\usepackage{subfigure}
\usepackage{amsmath}
\usepackage{amsfonts}
\usepackage{multirow}
\usepackage{epic}
\usepackage{listings}
\ifCLASSINFOpdf

\else

\fi

\begin{document}

\newtheorem{thm}{Theorem}
\newtheorem{cor}{Corollary}
\newtheorem{lem}{Lemma}
\unitlength=0.6mm

\title{Unsteady Reversed Stagnation-Point Flow over a Flat Plate}

\author{Vai~Kuong~Sin,~\IEEEmembership{Member,~ASME; Fellow,~MIEME,}
       and Chon Kit  Chio 
\thanks{V. K. Sin is with the Department
of Electromechanical Engineering, University of Macau, Macao SAR,
China, E-mail: (vksin@umac.mo).}% <-this % stops a space
%\thanks{Manuscript received March 27, 2012; revised April 02, 2012.}
}

\markboth{Journal of \LaTeX\ Class Files,~Vol.~6, No.~1, January~2007}%
{Shell \MakeLowercase{\textit{et al.}}: Bare Demo of IEEEtran.cls for Journals}



\maketitle
\thispagestyle{empty}

\begin{abstract}
%\boldmath
This paper investigates the nature of the development of two-dimensional laminar flow of an incompressible fluid at the reversed stagnation-point. ". In this study, we revisit the problem of reversed stagnation-point flow over a flat plate. Proudman and Johnson (1962) first studied the flow and obtained an asymptotic solution by neglecting the viscous terms. This is no true in neglecting the viscous terms within the total flow field. In particular it is pointed out that for a plate impulsively accelerated from rest to a constant velocity V0 that a similarity solution to the self-similar ODE is obtained which is noteworthy completely analytical.
\end{abstract}


% \fi
\IEEEpeerreviewmaketitle
\section{Introduction}
\IEEEPARstart{T}{he} full Navier-Stokes equations are difficult or impossible to obtain an exact solution in almost every real situation because of the analytic difficulties associated with the nonlinearity due to convective acceleration. The existence of exact solutions are fundamental not only in their own right as solutions of particular flows, but also are agreeable in accuracy checks for numerical solutions.

\begin{figure}[!htb]
\centering
\includegraphics[width=7cm]{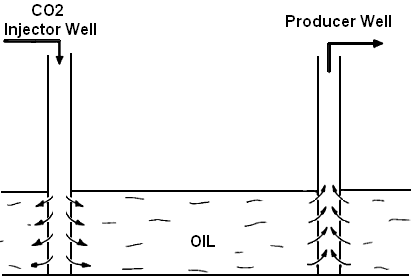}
\caption{Oil recovery industry }
\label{goil}
\end {figure}

In some simplified cases, such as a fluid travels through a rigid body (e.g., missile, sports ball, automobile, spaceflight vehicle), or in oil recovery industry crude oil that can be extracted from an oil field is achieved by gas injection, as shown in Fig.~(\ref{goil}), or equivalently, an external flow impinges on a stationary point called stagnation-point that is on the surface of a submerged body in a flow, of which the velocity at the surface of the submerged object is zero. A stagnation-point flow develops and the streamline is perpendicular to the surface of the rigid body. The flow in the vicinity of this stagnation point is characterized by Navier-Stokes equations. By introducing coordinate variable transformation, the number of independent variables is reduced by one or more. The governing equations can be simplified to the non-linear ordinary differential equations and are analytic solvable.  The classic problems of two-dimensional stagnation-point flows can be analyzed exactly by Hiemenz \cite{hiemenz1911grenzschicht}, one of Prandtl's first students. The result is an exact solution for flow directed perpendicular to an infinite flat plate. Howarth \cite{howarth1951cxliv} and Davey \cite{davey1961boundary} extended the two-dimensional and axisymmetric flows to three dimensions, which is based on boundary layer approximation in the direction normal to the plane.

\begin{figure}[!htb]
\centering
\includegraphics[width=8cm]{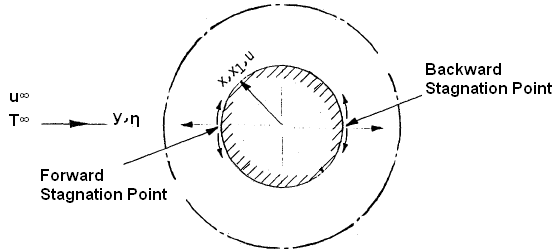}
\caption{Sketch of forward stagnation point and backward stagnation point}
\label{bStag}
\end {figure}

If the rear of the rigid body is not tapered, a stagnation-point flow also develops in the rear of the body, as shown in Fig.~(\ref{bStag}). The flow in the vicinity of this reversed stagnation point is governed by boundary-layer separation and vortex generation and the reverse stagnation point flow develops.  The main difference between these two flows is the change in the flow direction. Reversed stagnation-point flows against an infinite flat wall do not have analytic solution in two dimensions, but certain reverse flows have solution in three dimensions \cite{davey1961boundary}.

Proudman and Johnson \cite{proudman1962boundary} first suggested that the convection terms dominate in considering the inviscid equation in the body of the fluid. By introducing a very simple function of a particular similarity variable and neglecting the viscous forces away from the plane, they obtained an asymptotic solution in reversed stagnation-point flow, describing the development of the region of separated flow for large time $t$. In their solution, the phenomenon of separation is described near a plane that represents the rear-stagnation point of a cylinder is set in motion impulsively with a constant velocity normal to the surface of the plane. Robins and Howarth \cite{robins1972boundary} have recently extended the asymptotic solution, finding the higher order terms by singular perturbation methods. They indicated that the viscous forces cannot be ignored in the governing equation because of a consistent asymptotic expansion in both this outer inviscid region and also in the inner region near the plane. Smith \cite{smith1977development} generalized the solution of Proudman and Johnson with both viscous and convection terms in balance by considering the monotonic potential flow when the time is relatively large.

Numerical simulation of reversed stagnation-point flow with full Navier-Stokes equations has been studied in \cite{sin2011computational}. In the present study, the unsteady reversed stagnation-point flow is investigated. The flow is started impulsively in motion with a constant velocity away from near the stagnation point. A similarity solution of full Navier-Stokes equations is solved by applying numerical method.

\section{Flow Analysis Model}
\begin{figure}[htbp]
\begin{center}
\includegraphics[width= 6cm]
{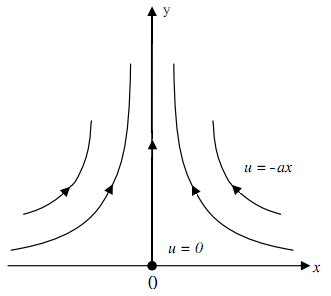}
\caption{Coordinate system of reversed stagnation-point flow}
\label{csp}
\end {center}
\end {figure}

The viscous fluid flows in a rectangular Cartesian coordinates $(x,y,z)$, Fig.~\ref{csp}, which illustrates the motion of external flow directly moves perpendicular out of an infinite flat plane wall. The origin is the so-called stagnation point and $z$ is the normal to the plane.

By conservation of mass principle with constant physical properties , the equation of continuity is:
\begin{equation}
\frac{\partial u}{\partial x}+\frac{\partial v}{\partial y}=0
\label{eq:e0}
\end{equation}
We consider the two-dimensional reversed stagnation-point flow in unsteady state and the flow is bounded by an infinite plane $y=0$, the fluid remains at rest when time $t<0$.  At $t=0$, it starts impulsively in motion which is determined by the stream function
\begin{equation}
\psi= -\alpha x y
\end{equation}

At large distances far above the planar boundary, the existence of the potential flow implies an inviscid boundary condition. It is given by

\begin{subequations}
\begin{gather}
u= -\alpha x\\
v=V_0
\end{gather}
\end{subequations}
where $u$ and $v$ are the components of flow velocity, $A$ is a constant proportional to $V_0/L$, $V_0$ is the external flow velocity removing from the plane and $L$ is the characteristic length. We have $u=0$ at $x=0$ and $v=0$ at $y=0$, but the no-slip boundary at wall $(y=0)$ cannot be satisfied.

For a viscid fluid the stream function, since the flow motion is determined by only two factors,  the kinematic viscosity $\nu$  and $\alpha$, we consider the following modified stream function
\begin{subequations}
\begin{gather}
\psi = -\sqrt{A\nu}xf(\eta, \tau)\label{psi1}\\
\eta = \sqrt{\frac{A}{\nu}}y\\
\tau = At
\end{gather}
\end{subequations}
where  $\eta$ is the non-dimensional distance from wall and $\tau$ is the non-dimensional time. Noting that the stream function automatically satisfies equation of continuity (\ref{eq:e0}) . The Navier-Stokes equations \cite{white-fluid} governing the unsteady flow with constant physical properties are
\begin{subequations}
 \begin{gather}
    \frac{\partial u}{\partial t}+ u \frac{\partial u}{\partial x}+v\frac{\partial u}{\partial y}=  -\frac{1}{
     \rho}\frac{\partial p}{\partial x}+ \nu (\frac{\partial^2 u}{\partial
     x^2}+\frac{\partial ^2u}{\partial y^2})
     \label{e3}\\
     \frac{\partial v}{\partial t}+ u \frac{\partial v}{\partial x}+v\frac{\partial v}{\partial y}=  -\frac{1}{
     \rho}\frac{\partial p}{\partial y}+ \nu (\frac{\partial^2 v}{\partial
     x^2}+\frac{\partial ^2v}{\partial y^2})
     \label{e4}
  \end{gather}
  \label{e3_4}
\end{subequations}
where $u$ and $v$ are the velocity components along $x$ and $y$ axes, and $\rho$ is the density.

Substituting $u$ and $v$ into the governing equations results a simplified partial differential equation. From the definition of stream function, we have
\begin{subequations}
  \begin{gather}
u= \frac{\partial \psi}{\partial y}  =  -Axf_{\eta}\\
v=-\frac{\partial \psi}{\partial x}
 =  \sqrt{A\nu}f
  \end{gather}
\end{subequations}

The governing equations can be simplified by a similarity transformation when several independent variables appear in specific combinations, in flow geometries involving infinite or semi-infinite surfaces. This leads to rescaling, or the introduction of dimensionless variables, converting the original systems of partial differential equations into a partial differential equation.
\begin{subequations}
  \begin{gather}
    -A^2xf_{\eta\tau}+A^2x(f_{\eta})^2-A^2xff_{\eta\eta}=   -\frac{1}{ \rho}
     \frac{\partial p}{\partial x}- A^2xf_{\eta\eta\eta}
    \label{e5}\\
     A\sqrt{A\nu}f_{\tau}+A\sqrt{A\nu}ff_{\eta}=   -\frac{1}{ \rho}
     \frac{\partial p}{\partial y}+ A\sqrt{A\nu} f_{\eta\eta}
    \label{e6}
  \end{gather}
\end{subequations}

The pressure gradient can be again reduced by a further differentiation Eq.~(\ref{e6}) with respect to $x$. That is
  \begin{equation}
\frac{\partial^2 p}{\partial x \partial y}=0
  \end{equation}
and Eq. (\ref{e5}) reduces to
  \begin{equation}
    [f_{\eta\tau}-(f_{\eta})^2+ff_{\eta\eta}-f_{\eta\eta\eta}]_{\eta}=0.
     \label{e7}
  \end{equation}
The initial and boundary conditions are
  \begin{subequations}
     \begin{gather}
        f(\eta, 0) \equiv \eta ~~~~~~~~~~~~~~~(\eta \neq 0)\\
        f(0,\tau)= f_{\eta}(0,\tau)=0~~(t\neq 0)\\
        f(\infty,\tau)\sim \eta~~~~~~~~~~~~~~~~~~~~~~
    \end{gather}
 \label{e8}
  \end{subequations}
The last condition reduces the above differential Eq.~(\ref{e7}) to the form
  \begin{equation}
    f_{\eta\tau}-(f_{\eta})^2+ff_{\eta\eta}-f_{\eta\eta\eta}=-1,
     \label{e9}
  \end{equation}
with the boundary conditions
  \begin{subequations}
     \begin{gather}
        f(0,\tau)= f_{\eta}(0,\tau)=0\\
        f_{\eta}(\infty,\tau) = 1.~~~~~~~~~~
    \end{gather}
 \label{e10}
  \end{subequations}
Eq.~(\ref{e9}) is the similarity equation of the full Navier-Stokes equations at two-dimension reversed stagnation point. The coordinates $x$ and $y$ are vanished, leaving only a dimensionless variable $\eta$.
Under the boundary conditions $f_{\eta}(\infty,\tau) = 1$, when the flow is in steady state such that $f_{\eta\tau}\equiv 0$, the differential equation has no solution.

\section{Similarity Analysis}

\subsection{Asymptotic solution}
When $\tau$ is relatively small, Proudman and Johnson \cite{proudman1962boundary} first considered  the early stages of the diffusion of the initial vortex sheet at $y=0$. They suggested that, when the flow is near the plane region, the viscous forces are dominant, and the viscous term in the governing Navier-Stokes equations is important only near the boundary.

On the contrary, the viscous forces were neglected away from the wall. The convection terms dominate the motion of external flow in considering the inviscid equation in the fluid. They considered the similarity of the inviscid equation

  \begin{equation}
    f_{\eta\tau}-(f_{\eta})^2+ff_{\eta\eta}+1=0.
     \label{e11}
  \end{equation}
Proudman and Johnson obtained a similarity solution of (\ref{e11}) is in the form
 \begin{equation}
f(\eta,\tau)=e^{\tau} F(\gamma)
\label{eq2_25}
  \end{equation}
\begin{equation}
F(\gamma)=\gamma -\frac{2}{c}(1-e^{-c\gamma} )
\label{e14}
\end{equation}
 where $\gamma=\displaystyle \frac{\eta}{\lambda(\tau)}$ and $c$ is a constant of integration. Robins and Howarth \cite{robins1972boundary} estimated the value of $c$ to be $3.51$. This solution describes the flow in the outer region, moving away from the plane with a constant velocity. It can be checked that the viscous term $f_{\eta\eta\eta}$ is still small compared to the convective terms, so that their assumsion of neglecting the viscous term is still valid.

In the inner region, the viscous term cannot be neglected and the no-slip condition must be satisfied on the wall. When $\tau \rightarrow \infty$ and $\eta /e^{\tau}$ is relatively small, the solution (\ref{e14}) yields
$$ F \sim -\gamma =-\eta e^{-\tau}$$
and
\begin{equation}
f = -\eta, ~~~~~f'=-1
\end{equation}
Substituting in (\ref{e9}) becomes
\begin{equation}
  \left\{
  \begin{array}{rr}
     f'''- ff''+(f')^2-1= 0 \\
     f(0)= f'(0)=0 \\
     f'(\infty) = -1
  \end{array}
  \right.
  \label{q:e12}
\end{equation}

\begin{figure}[htbp]
\begin{center}
\includegraphics[width= 7cm]
{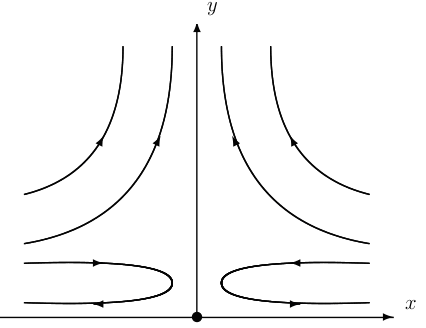}
\caption{Streamlines of reversed stagnation-point flow}
\label{vort}
\end {center}
\end {figure}

This is exactly the classic stagnation-point problem (Hiemenz \cite{hiemenz1911grenzschicht}) by changing the sign in $f$. It is a third-order nonlinear ordinary differential equation and does not have an analytic solution, and thus it is necessary to solve it numerically. The general features of the predicted streamline are sketched in Fig. (\ref{vort}).
\\

Although an asymptotic solution was obtained, it can easily been observed that this is no true in neglecting the viscous terms within the total flow field. No exact solutions in both outer and inner regions were discovered.

\subsection{Particular Solution}
In our two-dimensional model, the fluid remains at rest when time $t<0$ and is set in motion at $t>0$ such that at large distances far above the planar boundary the potential flow is a constant $V_0$ for all value of $t$. Both Proudman and Johnson \cite{proudman1962boundary}, and Robins and Howarth \cite{robins1972boundary} have set $V_0= 1$ and the corresponding boundary condition $f_{\eta}(\infty,\tau) = 1$.

When the flow is in steady state such that $f_{\eta\tau}\equiv 0$, it was proven that the similarity velocity $f_\eta(\eta)$ cannot ultimately approach to 1. The differential equation has no solution. In this chapter, if the potential flow $V_0$ is restricted not to be a constant, the boundary condition $f_{\eta}(\infty,\tau)$ results in a time dependent function and then we obtain another approach of similarity solution in reversed stagnation-point flow.

As with the governing equation of reversed stagnation-point flow, we can write the stream function as
\begin{subequations}
\begin{gather}
\psi = -\sqrt{A\nu}xf(\eta, \tau)\label{psi1}\\
\eta = \sqrt{\frac{A}{\nu}}y\\
\tau = At
\end{gather}
\end{subequations}
where $A$ is a constant proportional to $V_0(\tau)/L$, $V_0(\tau)$ is the external flow velocity removing from the plane and $L$ is the characteristic length. These result in the  gonvering equation (\ref{e7})
  \begin{equation}
    [f_{\eta\tau}-(f_{\eta})^2+ff_{\eta\eta}-f_{\eta\eta\eta}]_{\eta}=0.
     \label{e3_1}
  \end{equation}
After integration with respect to $\eta$, we have
\begin{equation}
f_{\eta\tau}-(f_{\eta})^2+ff_{\eta\eta}-f_{\eta\eta\eta}=-C(\tau),
\end{equation}
Under the boundary conditions $f_{\eta}(\infty,\tau) = 1$, the value of $C(\tau)$ should be a constant and equal to $1$.  If the boundary condition $f_{\eta}(\infty,\tau)$ is restricted not to be a constant,t, following the procedure of \cite{shapiro2006analytical}, a particular time-dependence function $C(\tau)$ may be expressed in the form
\begin{equation}
C(\tau)=\frac{c}{\tau^2}
\end{equation}
where $c$ is an arbitrary constant. The partial differential equation can be simplified by a similarity transformation when a new similarity variable is introduced. This converts the original partial differential equation into an ordinary differential equation. For a time dependent function, we introduce the diffusion variable transformation \cite{smith1977development}
\begin{subequations}
\begin{gather}
\varsigma  = \eta \sqrt{\frac{1}{\tau}} \\
f(\eta,\tau)=\frac{1}{\sqrt{\tau}}F(\varsigma)
\end{gather}
\end{subequations}
Here $\varsigma$ is the time combined nondimensional variable and $F(\varsigma )$ are the nondimensional velocity functions. Substitution of the similarity transformation yields an ordinary differential equation

  \begin{equation}
   -\frac{1}{2}\varsigma F''-F'-F'^2+FF''-F'''=-c
 \label{e3_6}
  \end{equation}

Equation (\ref{e3_6}) is a third-order nonlinear ordinary differential equation and a key step in obtaining an analytical solution is to rearrange the equation in an autonomous differential equation. In mathematics, an autonomous differential equation is a system of ordinary differential equations which does not explicitly depend on the independent variable.

In order to omitting the variable $\varsigma$ in the differential equation, it is recognized a change of variable
  \begin{equation}
  Q= F-\frac{1}{2}\varsigma
 \label{e3_0}
  \end{equation}
and the equation becomes to an autonomous differential equation
  \begin{equation}
QQ''-2Q'-Q'^2-Q'''=-c+\frac{3}{4}
 \label{e3_9}
  \end{equation}
In our analysis, $P=Q'$ is the dependent variable and $Q$ is the independent variable. Equation (\ref{e3_9}) is rearranged as
  \begin{equation}
QP'-2P-P^2-P''=-c+\frac{3}{4}
 \label{e3_7}
  \end{equation}
and the chain rule reduces equation (\ref{e3_7}) to a second-order ordinary differential equation
\begin{equation}
QP\frac{dP}{dQ}-2P-P^2-P\frac{d}{dQ}\left(P\frac{dP}{dQ}\right)=-c+\frac{3}{4}
 \label{e3_7a}
  \end{equation}
Equation (\ref{e3_7a}) is analytically solvable that the solution might be expressed as a low order polynomial. It is suggested that
  \begin{equation}
P=a+bQ+dQ^2
 \label{e3_8}
  \end{equation}
and substituting into equation (\ref{e3_7}) and comparing the coefficients in the powers of $Q$ results in a system of linear algebraic equation
\begin{subequations}
\begin{gather}
2a^2d+ab^2+2a-a^2=-c+\frac{3}{4}\\
8abd+b^3+2b+ab=0\\
8ad^2+(7b^2+2)d=0\\
12bd^2-bd=0\\
6d^3-d^2=0
\end{gather}
\end{subequations}
Solving the related algebraic equation, we have
  \begin{equation}
a=-\frac{3}{2}, ~~~~b=0, ~~~~c=\frac{3}{4},~~~~d=\frac{1}{6}
  \end{equation}
Substituting the constant into equation (\ref{e3_8}) yields the first-order differential equation
  \begin{equation}
Q'=-\frac{3}{2}+\frac{1}{6}Q^2
 \label{e3_15}
  \end{equation}
Equation (\ref{e3_15}) is Riccati equation , which is any ordinary differential equation that is quadratic in the unknown function. The standard form of Riccati equation is
  \begin{equation}
Q'=RQ^2+SQ+T
  \end{equation}
The solution of Riccati equation can be obtained by a change of dependent variable, where the dependent variable $y$ is changed to $q$ by \cite{zwillinger1998handbook}
  \begin{equation}
Q=-\frac{q'}{q}\frac{1}{R}
  \end{equation}
By identifying $R=\frac{1}{6}$, $S=0$ and $T=-\frac{3}{2}$, the change of variables in equation (\ref{e3_15}) becomes
  \begin{equation}
Q=-\frac{q'}{\frac{1}{6}q}=-\frac{6 q'}{q}
 \label{e3_16}
  \end{equation}
so the equation (\ref{e3_15}) becomes an second-order linear differential equation
  \begin{equation}
q''-\frac{1}{4}=0
  \end{equation}
The general solution to this equation  is
  \begin{equation}
q=A~cosh\frac{\varsigma}{2}+B~sinh\frac{\varsigma}{2}
  \end{equation}
where $A$ and $B$ are arbitrary constants. Applying this solution in equation (\ref{e3_0}) leads to the general solution of equation (\ref{e3_6})
  \begin{equation}
F(\varsigma)=\frac{\varsigma}{2}-\frac{3A~sinh\frac{\varsigma}{2}+3B~cosh\frac{\varsigma}{2}}{A~cosh\frac{\varsigma}{2}+B~sinh\frac{\varsigma}{2}}
 \label{e3_21}
  \end{equation}
Application of the no-slip condition $F(0)=0$ leads to the determination of the constant $B=0$, so the exact solution becomes
  \begin{equation}
F(\varsigma)=\frac{\varsigma}{2}-3~tanh\frac{\varsigma}{2}
 \label{e3_22}
  \end{equation}
Collecting results, the velocity functions become
  \begin{equation}
f(\eta,\tau)=\frac{1}{\sqrt{\tau}}\left(\frac{\varsigma}{2}-3~tanh\frac{\varsigma}{2}\right)
 \label{e3_23}
  \end{equation}
where $\varsigma=\displaystyle \sqrt{\frac{A}{\nu\tau}}y$ is the non-dimensional distance from the plate. In view of (\ref{e3_23}), the flow far form the boundary $(x, \varsigma \rightarrow \infty)$ becomes
  \begin{equation}
f(\eta,\tau) \rightarrow \frac{1}{\sqrt{\tau}}\left(\frac{\varsigma}{2}-3\right)
 \label{e3_24}
  \end{equation}

We obtain a particular solution of the unsteady reversed stagnation-point flow. The above solution is obtained in the similarity framework for unsteady viscous flows. The appearance of this positive factor in the first terms of (\ref{e3_24}) shows that this remote flow is directed toward the axis of symmetry and away from the plate. The second term in (\ref{e3_24}) describes a uniform current directed toward the plate. An adverse pressure gradient near the wall region leads to boundary-layer separation and associated flow reversal.

The particular solution is noteworthy in that it is completely analytical, but it is limited to the region far away from the wall in the presence of non-zero term $F'(0)=-1$. The no-slip boundary condition is not satisfied completely near the wall region.

\subsection{Numerical Solution}
Since the analytical solution does not satisfy the no-slip condition $F'=0$, it is convenient  to solve the similarity equation numerically. The similarity equation and the relevant boundary conditions are
\begin{equation}
  \left\{
  \begin{array}{rr}
      -\frac{1}{2}\varsigma F''-F'-F'^2+FF''-F'''=-c \\
     F(0)= F'(0)=0 \\
     F'(\infty) = \frac{1}{2}
  \end{array}
  \right.
  \label{q:e12}
\end{equation}
where $c=\frac{3}{4}$ in order to satisfy the unsteady viscous flows in the outer region.

This equation is a third-order nonlinear ordinary differential equation. It is convenient when solving an ODE system numerically to describe the problem in terms of a system of first-order equations in MATLAB\cite{shampine2003solving}.

For example when solving an $n^{th}$-order problem numerically is common practice to reduce the equation to a system of $n$ first-order equations. Then, by defining
$y_1 = F,~y_2 = F',~y_3 = F''$, the ODE reduces to the form
    \begin{equation}
\frac{d\mathbf{y}}{dx} =
\begin{bmatrix}
  y_2  \\
  y_3  \\
  c-\frac{1}{2}\varsigma y_3-y_2-y_2^2 +y_1*y_3
\end{bmatrix}
    \label {eq:e22}
\end {equation}

The first task is to reduce the equation above to a system of first-order equations and define in MATLAB a function to return these. The relevant MATLAB expression for Eq.~(\ref{eq:e22}) would be:

\begin{lstlisting}[label=MATLAB,caption=System of first-order equations ]
function dy = stagnation(t,y)
c=3/4;
dy = zeros(3,1);
dy(1) = y(2);
dy(2) = y(3);
dy(3) = c-1/2*t*y(3)-y(2)-y(2)*y(2)+y(1)*y(3);
\end{lstlisting}

Later, we need to change the boundary value into initial value, because $ode45$, a ode solver in MATLAB, can only solve the initial value problem. From Eq.~(\ref{q:e12}), we need to gauss the value of $F''(0)$ such that $F'(\infty) = \frac{1}{2}$. The commands written in MATLAB would be

\begin{lstlisting}[label=MATLAB,caption=ODE solver]
function main
x=20;
[T,Y] = ode45(@stagnation,[0 x],[0 0 -1.54306]);
plot(T,Y(:,1),'-',T,Y(:,2),'--',T,Y(:,3),'-.')
\end{lstlisting}

\begin{figure}[!htbp]
\begin{center}
\includegraphics[width= 9cm]
{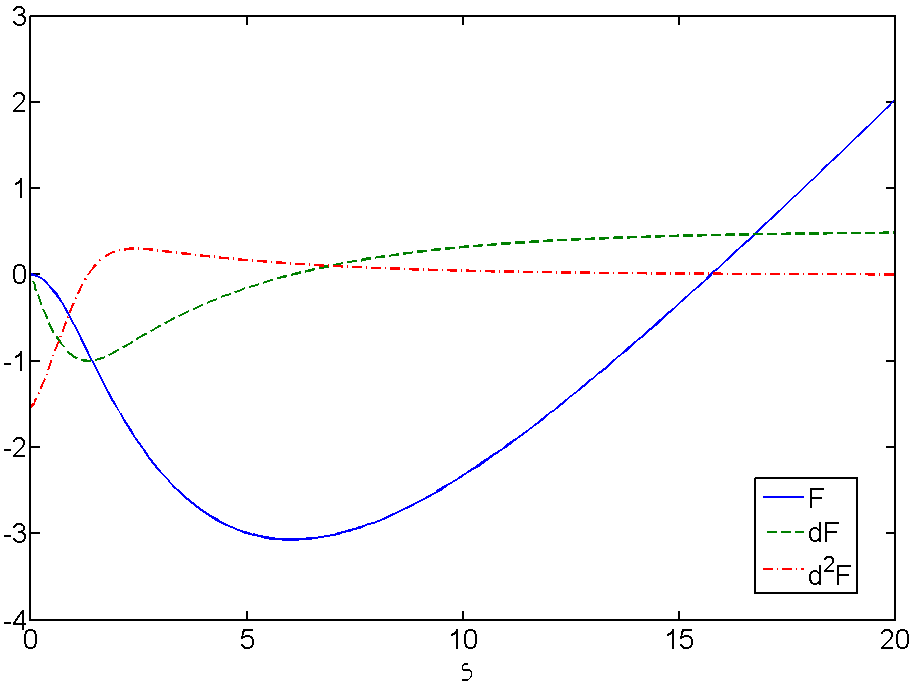}
\caption{Numerical solutions of reversed stagnation-point flow}
\label{g1}
\end{center}
\end{figure}

The numerical solution for two-dimensional stagnation-point  flow is shown in Fig.(\ref{g1}).

\begin{figure}[!htb]
\centering
\includegraphics[width=8.5cm]{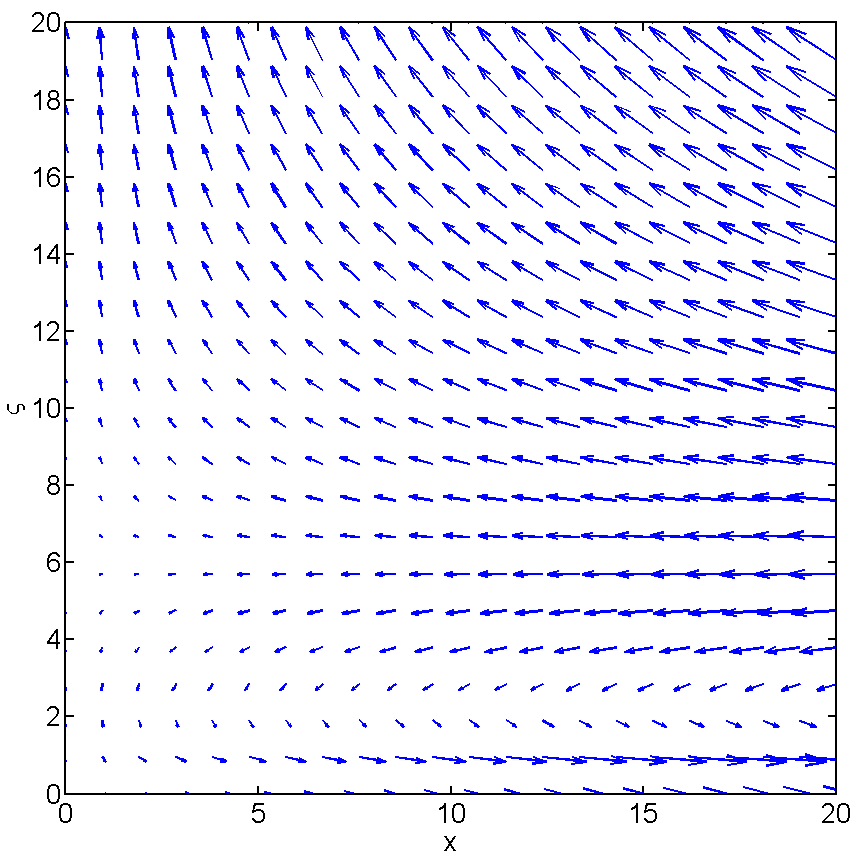}
\caption{Similarity velocity field as a function of $\varsigma$}
\label{stream}
\end {figure}

This solution is a similarity solution of the reversed stagnation-point flow over a flat plate. It describes an unsteady viscous flow in both outer and inner regions.  A single dividing streamline plane separates streamlines approaching the plate from external flow streamlines. Though the viscous term $F'''$ is still small compared to the convective terms in the outer region, this is no true in neglecting the viscous terms within the total flow field. The similarity velocity field is show in Fig.(\ref{stream}).

\section{Conclusion}
The foregoing study constitutes a similarity solution of the unsteady Navier-Stokes equations for reversed stagnation-point flow in the idealized case of an infinite plane boundary. In order to analyze the flow for non-zero values of $x$, it is required to convert the full Navier-Stokes equations. This problem is now being studied by applying numerical method. The solution is obtained in the classical similarity framework for unsteady viscous flows.

In the case of numerical methods, a brief analysis of the new solution is discussed. When the flow was near the plane wall region, the viscous forces were dominant, and the viscous term in the governing Navier-Stokes equations was important only near the boundary. On the contrary, the viscous forces were negligible when they were away from the wall.

\section*{Acknowledgement}
The research was partially supported by Research Committee of University of Macau through the grants RG079/09-10S//11T/SVK/FST, and by Science and Technology Development Fund (FDCT) of Macao SAR through the grant 034/2009/A.

\bibliographystyle{ieeetr}	
\bibliography{myrefs}

\begin{thebibliography}{10}

\bibitem{hiemenz1911grenzschicht}
K.~Hiemenz, ``{Die Grenzschicht an einem in den gleichf{\"o}rmigen
  Fl{\"u}ssigkeitsstrom eingetauchten geraden Kreiszylinder, Dingl.
  Polytech},'' {\em J}, vol.~326, pp.~321--410, 1911.

\bibitem{howarth1951cxliv}
L.~Howarth, ``{CXLIV. The boundary layer in three dimensional flow.-Part II.
  The flow near a stagnation point},'' {\em Philosophical Magazine (Series 7)},
  vol.~42, no.~335, pp.~1433--1440, 1951.

\bibitem{davey1961boundary}
A.~Davey, ``{Boundary-layer flow at a saddle point of attachment},'' {\em
  Journal of Fluid Mechanics}, vol.~10, pp.~593--610, 1961.

\bibitem{proudman1962boundary}
I.~Proudman and K.~Johnson, ``{Boundary-layer growth near a rear stagnation
  point},'' {\em Journal of Fluid Mechanics}, vol.~12, no.~02, pp.~161--168,
  1962.

\bibitem{robins1972boundary}
A.~Robins and J.~Howarth, ``{Boundary-layer development at a two-dimensional
  rear stagnation point},'' {\em Journal of Fluid Mechanics}, vol.~56, no.~01,
  pp.~161--171, 1972.

\bibitem{smith1977development}
S.~Smith, ``The development of the boundary layer at a rear stagnation point,''
  {\em Journal of Engineering Mathematics}, vol.~11, no.~2, pp.~139--144, 1977.

\bibitem{sin2011computational}
V.~K. Sin and C.~K. Chio, {\em Computation of Non-Isothermal Reversed
  Stagnation-Point Flow over a Flat Plate}, ch.~Computational Simulations and
  Applications, pp.~159--174.
\newblock InTech, 2011.
\newblock ISBN: 978-953-307-430-6.

\bibitem{white-fluid}
F.~White, ``{Fluid Mechanics. 5th edt},'' 2003.

\bibitem{shapiro2006analytical}
A.~Shapiro, ``An analytical solution of the navier-stokes equations for
  unsteady backward stagnation-point flow with injection or suction,'' {\em
  ZAMM-Journal of Applied Mathematics and Mechanics/Zeitschrift f{\"u}r
  Angewandte Mathematik und Mechanik}, vol.~86, no.~4, pp.~281--290, 2006.

\bibitem{zwillinger1998handbook}
D.~Zwillinger, {\em Handbook of differential equations}.
\newblock Academic Press, 1998.

\bibitem{shampine2003solving}
L.~Shampine, I.~Gladwell, and S.~Thompson, {\em {Solving ODEs with MATLAB}}.
\newblock Cambridge Univ Pr, 2003.

\end{thebibliography}

%\begin{biography}[{\includegraphics[width=1in,height=1.25in,clip,keepaspectratio]{mshell}}]{Michael Shell}

\begin{IEEEbiography}[{\includegraphics[width=1in,height=1.5in,clip,keepaspectratio]{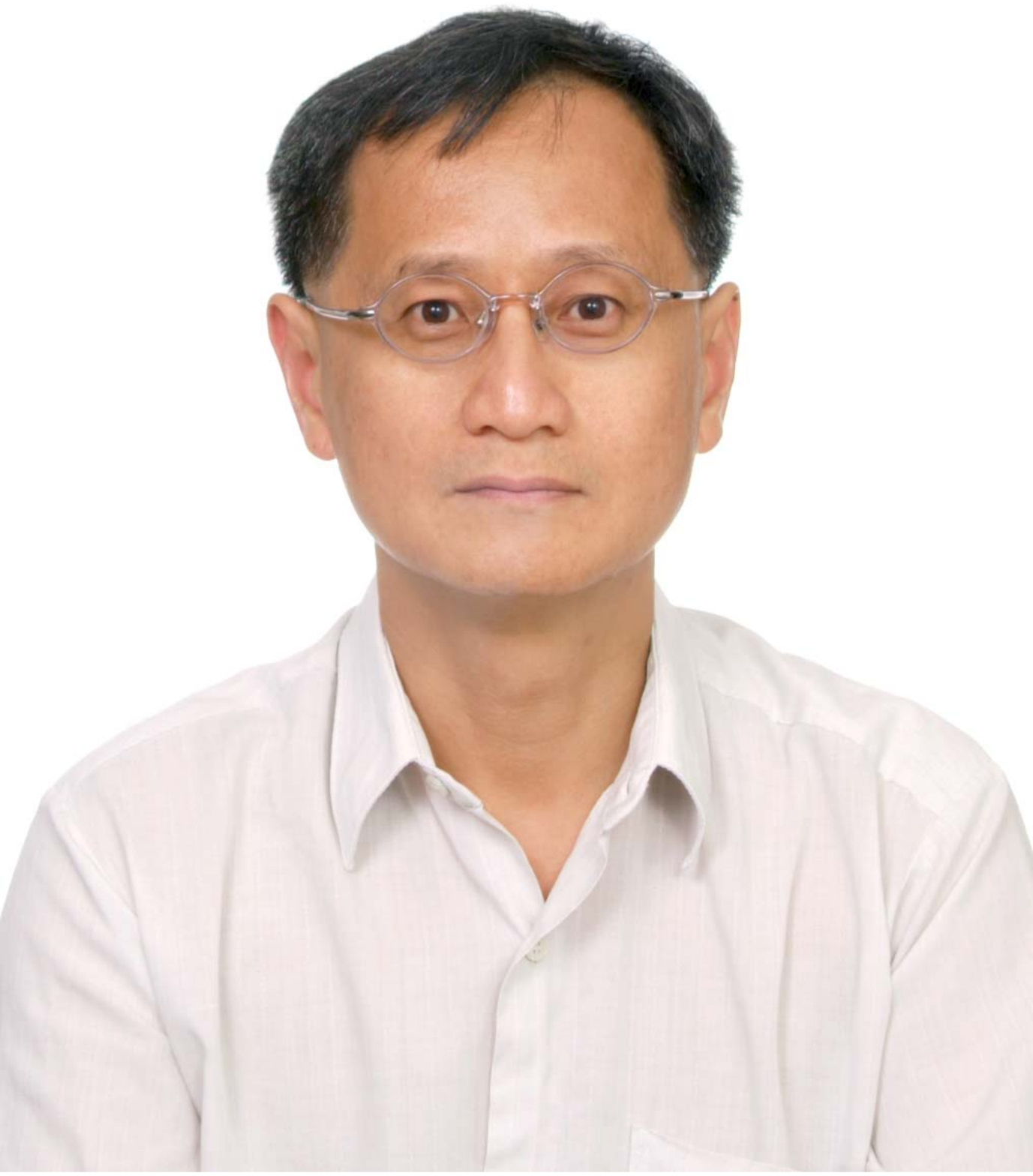}}]{Vai Kuong Sin}
is with the Department of Electromechanical Engineering, University
of Macau, Macao SAR, China, E-mail: vksin@umac.mo. His current
research interests include scientific computing, numerical
simulation, and computational fluid dynamics.
\end{IEEEbiography}

\begin{IEEEbiography}[{\includegraphics[width=1in,height=1.1in,clip,keepaspectratio]{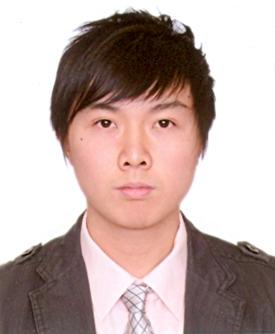}}]{Chio Chon Kit}
is with the Department of Electromechanical Engineering, University
of Macau, Macao SAR, China, E-mail: mb054275@umac.mo.
\end{IEEEbiography}


%\vfill


\end{document}